# Comparison and calibration of MP2RAGE quantitative T1 values to multi-TI inversion recovery T1 values

Adam M. Saunders,[a][*] Michael E. Kim,[b] Chenyu Gao,[a] Lucas W. Remedios,[b] Aravind R. Krishnan,[a] Kurt G. Schilling,[c-d] Kristin P. O'Grady,[c-e] Seth A. Smith,[c-e] Bennett A. Landman[a-e]

[a]Department of Electrical and Computer Engineering, Vanderbilt University, Nashville, TN, United States
[b]Department of Computer Science, Vanderbilt University, Nashville, TN, United States
[c]Vanderbilt University Institute of Imaging Science, Vanderbilt University Medical Center, Nashville, TN, United States
[d]Department of Radiology and Radiological Sciences, Vanderbilt University Medical Center, Nashville, TN, United States
[e]Department of Biomedical Engineering, Vanderbilt University, Nashville, TN, United States

*adam.m.saunders@vanderbilt.edu

**Acknowledgements**

This work was supported by NIH grants 1R01EB017230, 5K01EB030039, 5F32NS101788, K01EB032898, and National Cancer Institute R01 CA253923. This work was supported by Integrated Training in Engineering and Diabetes, grant number T32 DK101003. This work was conducted in part using the resources of the Center for Human Imaging at the Vanderbilt University Institute of Imaging Science (NIH OD award number 1S10OD030389-01) and the Advanced Computing Center for Research and Education at Vanderbilt University, Nashville, TN. The Vanderbilt Institute for Clinical and Translational Research (VICTR) is funded by the National Center for Advancing Translational Sciences (NCATS) Clinical Translational Science Award (CTSA) Program, Award Number 5UL1TR002243-03. The content is solely the responsibility of the authors and does not necessarily represent the official views of the NIH.

**Abstract**

While typical qualitative T1-weighted magnetic resonance images reflect scanner and protocol differences, quantitative T1 mapping aims to measure T1 independent of these effects. Changes in T1 in the brain reflect structural changes in brain tissue. Magnetization-prepared two rapid acquisition gradient echo (MP2RAGE) is an acquisition protocol that allows for efficient T1 mapping with a much lower scan time per slab compared to multi-TI inversion recovery (IR) protocols. We collect and register B1-corrected MP2RAGE acquisitions with an additional inversion time (MP3RAGE) alongside multi-TI selective inversion recovery acquisitions for four subjects. We use a maximum a posteriori (MAP) T1 estimation method for both MP2RAGE and compare to typical point estimate MP2RAGE T1 mapping, finding no bias from MAP MP2RAGE but a sensitivity to B1 inhomogeneities with MAP MP3RAGE. We demonstrate a tissue-dependent bias between MAP MP2RAGE T1 estimates and the multi-TI inversion recovery T1 values. To correct this bias, we train a patch-based ResNet-18 to calibrate the MAP MP2RAGE T1 estimates to the multi-TI IR T1 values. Across four folds, our network reduces the RMSE significantly (white matter: from $0.30 \pm 0.01$ seconds to $0.11 \pm 0.02$ seconds, subcortical gray matter: from $0.26 \pm 0.02$ seconds to $0.10 \pm 0.02$ seconds, cortical gray matter: from $0.36 \pm 0.02$ seconds to $0.17 \pm 0.03$ seconds). Using limited paired training data from both sequences, we can reduce the error between quantitative imaging methods and calibrate to one of the protocols with a neural network.

## 1. Introduction

Typical structural magnetic resonance imaging (MRI) produces a qualitative image where the voxel values are weighted by the magnetic resonance tissue relaxation parameters. Anatomical information, scanner differences and acquisition protocol differences affect the voxel values in structural MRI [1]. In contrast, quantitative MRI aims to produce maps of specific parameters, such as T1, T2, and proton density, with voxels representing the quantitative value instead of arbitrary units, designed to provide a reproducible, standardized metric [1]. Quantitative MRI can provide the ability to infer histological information noninvasively [1,2].

Quantitative imaging of the longitudinal relaxation parameter T1 is often referred to as T1 mapping. Changes in T1 values reflect structural changes across pathologies [1,3,4]. There are several methods for quantitative T1 mapping. A standard family of methods are inversion recovery (IR) methods that sample the signal recovery curve at multiple points after an inversion pulse, like inversion recovery spin echo (IR-SE), although acquisition times are extremely long [5]. An alternative with a shorter acquisition time is echo planar imaging (IR-EPI) [6]. IR-EPI can suffer from spatial distortions, but Sanchez-Panchuelo et al. have proposed modifications to reduce the distortions and efficiently acquire multi-slice images [7]. Selective inversion recovery (SIR) is another multi-TI inversion recovery method that can produce quantitative T1 maps in addition to other metrics that quantify the magnetization transfer effect [8,9].

An alternative to multi-TI IR T1 mapping methods with many samples along the signal recovery curve, magnetization prepared two rapid acquisition gradient echoes (MP2RAGE) allows for high-resolution and efficient quantitative T1 mapping with a much shorter acquisition time than multi-TI IR methods [10]. Quantitative T1 mapping from an MP2RAGE image involves obtaining two gradient echo readouts (GREs) formed from a train of gradient echoes and combining them to form a T1-weighted image. This T1-weighted image is independent of effects from T2* and the static magnetic field $B_0$, though there are still higher-order effects from the radiofrequency field, referred to as $B_1^+$. $B_1^+$ inhomogeneities also affect multi-TI IR acquisitions. While we can optimize the acquisition parameters for MP2RAGE to minimize effects from $B_1^+$, there still may be effects from $B_1^+$ in the image and corresponding quantitative T1 map, and quantitative T1 mapping is sensitive to the values of $B_1^+$ and acquisition parameters (see supplementary materials). Eggenschwiler et al. have proposed modifications to MP2RAGE to account for inhomogeneities in $B_1^+$, where a second acquisition with a similar sequence called saturation-prepared with 2 rapid gradient echoes (Sa2RAGE) can provide iterative estimates for T1 and $B_1^+$ together [11]. Quantitative T1 mapping with MP2RAGE at 7T shows high reproducibility between sites [12].

Due to the long repetition time MP2RAGE$_{TR}$ between inversion pulses, modifications to the MP2RAGE sequence can allow for collecting multiple GREs beyond two GREs collected a typical MP2RAGE acquisition. Multi-echo MP2RAGE can be used to generate simultaneous quantitative mappings of T1, T2*, and quantitative susceptibility mapping [13–15]. The MPnRAGE sequence is another modification that uses a sliding window to effectively collect $n$ images with different T1 weighting contrasts [16]. MP3RAGE collects a third GRE with a high flip angle to generate T1 maps and $B_1^+$ maps simultaneously with analytical solutions for T1 and $B_1^+$ [17]. Another method also referred to as MP3RAGE uses the third GRE to simultaneously solve for T1 and inversion pulse efficiency [18].

Marques et al. have validated quantitative T1 mapping with MP2RAGE in phantoms and in vivo with comparisons to values found in the literature [10]. Gochberg and Gore have validated quantitative T1 mapping with SIR in phantoms and in vivo with animals [19], while Bagnato et al. and Cronin et al. have studied SIR T1 maps in vivo with human brains [8,20].

Previous studies have compared co-registered quantitative T1 values from MP2RAGE to several different multi-TI IR protocols [5,7,21]. Rioux et al. found that there are significant differences between MP2RAGE quantitative T1 values and inversion recovery fast spin echo (IR-FSE) quantitative T1 values, potentially because a biexponential fitting for IR-FSE models the inversion recovery well to account for magnetization transfer effects, but MP2RAGE uses a monoexponential model [21]. With only two to three GREs for MP2RAGE and MP3RAGE, we are limited in the number of samples we can use to fit MP2RAGE inversion recovery to a biexponential model. The longer inversion times for MP2RAGE can mitigate this effect but do not remove it completely [21].

Here, we aim to not only compare quantitative T1 values from MP2RAGE and multi-TI IR but also calibrate the MP2RAGE T1 values to the multi-TI IR T1 values using a deep learning model. Calibration between the two methods would allow us to collect MP2RAGE scans with low scan times per slab and calibrate the T1 values to IR T1 values, which has a much higher scan time per slice. Additionally, we use a maximum a posteriori (MAP) approach to MP2RAGE quantitative T1 mapping that allows for T1 estimation from MP2RAGE and MP3RAGE acquisitions and for the additional estimation of the standard deviation of T1 as an uncertainty estimate. We contribute two findings. First, we compare typical point estimate MP2RAGE quantitative T1 values to MAP T1 estimates that allow us to map T1 for both MP2RAGE and MP3RAGE and generate additional metrics like the standard deviation of T1, and we find no substantial bias between the MAP MP2RAGE and point estimate MP2RAGE, though MAP MP3RAGE is more sensitive to B1 inhomogeneities. Second, we find a substantial tissue-dependent bias between MAP MP2RAGE T1 estimates and multi-TI IR T1 values, and we correct this bias by calibrating the MAP MP2RAGE T1 estimates to the multi-TI IR T1 values using a deep learning model. The calibration model allows retrospective calibration of the T1 estimates with limited paired data.

## 2. Methods

### 2.1. Data acquisition and processing

To compare MP2RAGE T1 maps with multi-TI IR T1 maps, we obtained data from four subjects on a 7T Philips Achieva scanner. We obtained informed consent after explaining the procedure under a protocol approved by the local institutional review board. We collected a T1-weighted magnetization-prepared rapid acquisition gradient echo (MPRAGE) acquisition, a multi-TI IR acquisition, an MP2RAGE acquisition with an additional GRE (also known as an MP3RAGE acquisition), as well as a dual-TR actual flip angle acquisition for B1 correction of the MP2RAGE/MP3RAGE and multi-TI IR T1 maps. The radiofrequency hardware was an 8-channel transmit and multichannel receive coil with capability for head imaging.

The resolution of the MP2RAGE acquisition was 0.7 mm × 0.43 mm × 0.43 mm, while the resolution of the multi-TI IR acquisition was 0.86 mm × 0.86 mm × 2 mm. The dual-TR actual flip angle acquisition was a single axial slice of 10 mm (resolution 0.86 mm × 0.86 mm × 10 mm) with flip angles of 60 degrees. The field of view was 220 mm, and we used SENSE for encoding acceleration with a factor of 2. We registered the B1 map calculated online from the scanner to the multi-TI IR image. Note that this assumes the B1 inhomogeneities are equivalent along the axial slices of the multi-TI IR image.

For greater flexibility in T1 mapping while maintaining the same $MP2RAGE_{TR}$, we acquired three GREs for an MP3RAGE acquisition. The acquisition consisted of a typical MP2RAGE sequence modified by the addition of a third GRE. Because of the long $MP2RAGE_{TR}$, the MP3RAGE acquisition with two GREs was the same duration as an MP2RAGE acquisition with two GREs. The acquisition parameters were inversion times $TI_1 = 1010$ ms, $TI_2 = 3683$ ms, $TI_3 = 6355$ ms, a repetition time between excitation pulses of $TR = 6$ ms, and a repetition time between inversion pulses of $MP2RAGE_{TR} = 8.25$ s. The slab orientation was sagittal, with a field of view of 220 mm. We used SENSE for encoding acceleration with a factor of 2. The excitation pulse was a sinc-Gauss shape with a duration of 0.67 ms. Each GRE block used 225 excitation pulses and flip angles of 4°. The inversion pulse was a hyperbolic secant shape with an inversion pulse efficiency of 0.84. The inversion pulse efficiency was calculated from a simulation similar to the one described in Marques et al. [10]. We tested inversion pulse efficiency values of 0.84 and 0.96 as in [10] and found that the quantitative T1 values were largely insensitive to this value. The scan time was approximately 9 minutes for 206 sagittal slabs of 0.7 mm thickness. We used dcm2niix to convert DICOMs to NIFTI format, applying the Philips volume scaling to produce the precise floating point (i.e., not raw integer) representation. From the MP3RAGE acquisition, we also used the first two GREs to form a typical MP2RAGE image for comparison.

### 2.1.1. Segmentation and registration to multi-TI IR images

For comparing the errors in each tissue type, we used U-shaped medical image segmentation model with nested transformers (UNesT) to segment the image. UNesT is a state-of-the-art deep learning-based image segmentation model with pretrained weights for segmentation of brain tissue [22]. We combined the labels using a 12-level hierarchical model whose labels were selected by a neuroimaging expert [23]. We used the level with 8 clusters and analyzed clusters 3 (subcortical gray matter, SGM), 4 (cortical gray matter, CGM), and 7 (cerebral white matter, WM). Cluster 4 also included the amygdala and hippocampus, but these structures were not present in the mid-brain slices for which we collected multi-TI IR images. We eroded the CGM cluster mask with a cube with sides of length 3 voxels to mitigate partial volume effects from the cerebrospinal fluid (CSF). The other clusters were either background, CSF in the ventricles, or labels that were not present within the slices we analyzed. Next, we used the MPRAGE images to create skull-stripped masks using SynthStrip [24].

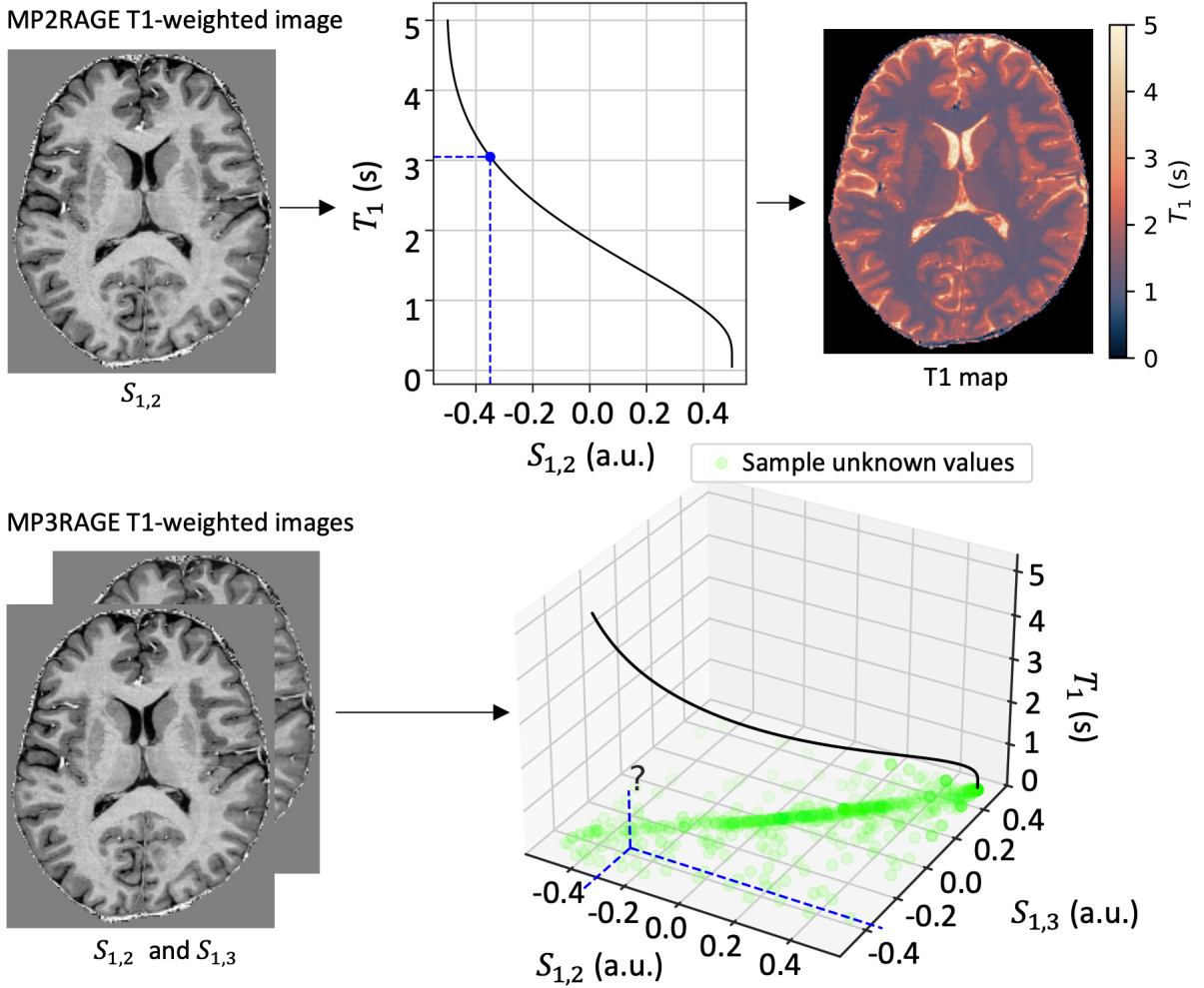

**Fig. 1.** Quantitative T1 mapping with MP2RAGE consists of creating a T1-weighted MP2RAGE image $S_{1,2}$ and then using the MP2RAGE model to find corresponding values of T1. However, this method provides no measures of uncertainty and does not extend to MP3RAGE, where $S_{1,2}$ may correspond to a different T1 value than $S_{1,3}$ using the MP2RAGE model. Therefore, the model is not directly invertible at all points. We visualize the T1 map using the color mapping recommended by Fuderer et al. [42].

To compare the images within a common spatial framework, we registered all images to the multi-TI IR image space using a rigid registration from the ANTs registration toolkit (version 2.4.4) [25]. Forming the T1-weighted MP2RAGE image from the two GREs involves dividing by the magnitudes of the GREs. When these values are near zero, the intensity is unstable, especially in the background. O'Brien et al. suggested adding a constant value $\beta$ to the denominator to create a more robust T1-weighted image for segmentation and registration with background noise suppressed [26]. Here, we used a constant $\beta$ value of 0.25 times the mean sum of the squared magnitudes of the GREs to register the MP2RAGE images to the multi-TI IR images. We used linear interpolation to resample to the coarser resolution of the multi-TI IR image, using a cube with sides of three voxel widths to erode the CGM label to account for partial volume effects from the CSF.

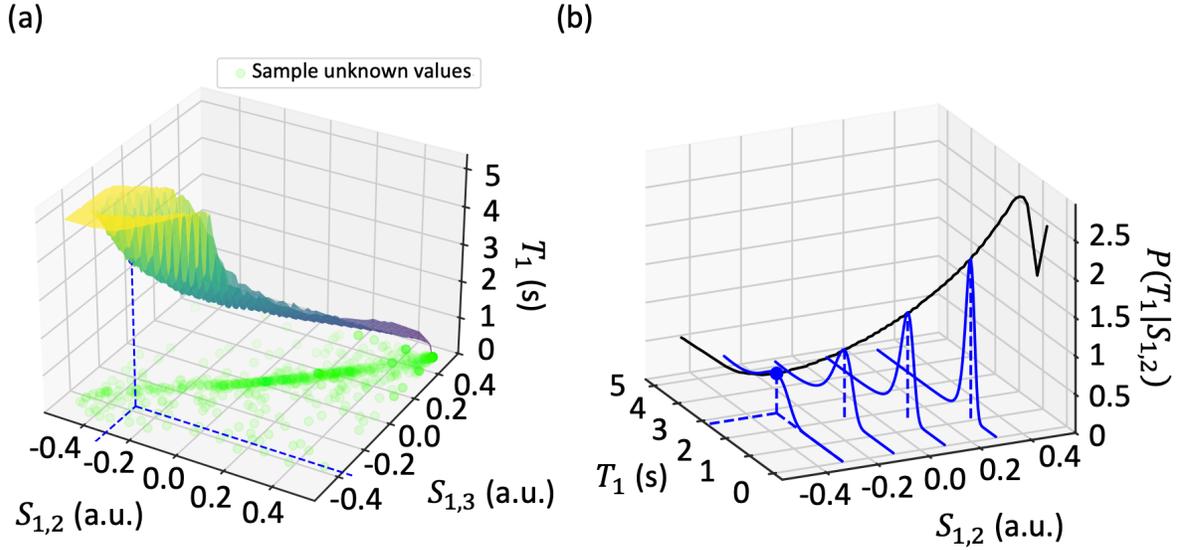

**Fig. 2.** To extend T1 mapping from MP2RAGE with a single T1-weighted image to MP3RAGE with multiple T1-weighted images (a) and allow for uncertainty estimation, we aim to find the posterior distribution of T1 given $S_{1,2}$ and $S_{1,3}$ (b). Using a Monte Carlo simulation to generate the posterior distribution, we can provide statistical measures such as the MAP estimate or standard deviation of T1.

## 2.2. MAP MP2RAGE and MAP MP3RAGE T1 mapping

To create MP2RAGE T1 maps, we follow the original point estimate method from Marques et al. as a baseline using the first two GREs [10]. To summarize, the method involves acquiring two complex-valued GREs at inversion times $TI_1$ and $TI_2$. The two GREs form the MP2RAGE image $S_{1,2}$ as

$$S_{1,2} = \mathrm{Re}\left(\frac{GRE_{TI_1}^* \, GRE_{TI_2}}{\left|GRE_{TI_1}\right|^2 + \left|GRE_{TI_2}\right|^2}\right),$$

(1)

where $*$ indicates the complex conjugate. From the acquisition parameters and the MP2RAGE model [10], we can create a lookup table and calculate what the image $S_{1,2}$ should be given a set of T1 values. We map T1 for the acquired image by finding the corresponding T1 value for a given MP2RAGE image value at each voxel.

While the MP2RAGE sequence aims to remove inhomogeneities due to the $B_1^+$ field by optimizing the sequence parameters [10], there are still $B_1^+$ biases present in the images. We must correct these biases to accurately compare MP2RAGE T1 maps with inversion recovery T1 maps [27]. Eggenschwiler et al. and Marques et al. used $B_1^+$ maps generated from saturation-prepared two rapid acquisition gradient echo (Sa2RAGE) in tandem with MP2RAGE images to correct for these biases [11,28]. For B1 correction of the MP2RAGE T1 maps, we used a B1 map calculated online from the scanner from the dual-TR actual flip angle acquisition [29]. We multiplied the B1 correction factor for each voxel by the nominal MP2RAGE flip angles when estimating the lookup table for the T1 maps. The derivation of the MP2RAGE model by Marques et al. takes into account the efficiency for the initial inversion pulse, but there are still higher order effects of $B_1^+$ inhomogeneities in the radiofrequency flip angles in the GRE block. For the MP3RAGE acquisition, we could choose to treat the inversion pulse efficiency as a free parameter and fit it from the data. Because we found that the quantitative T1 values were largely insensitive to the inversion pulse efficiency, we chose to fix this value and use a separate acquisition to find the $B_1^+$ correction factor a priori.

The third GRE in our MP3RAGE acquisition allows us to form two T1-weighted images from pairs of the GREs, $S_{1,2}$ and $S_{1,3}$. When extending from a single T1-weighted MP2RAGE image $S_{1,2}$ to MP3RAGE images $S_{1,2}$ and $S_{1,3}$, generating a T1 map is not straightforward. Due to variability inherent in the image acquisition, $S_{1,2}$ may correspond to a different T1 value than $S_{1,3}$, meaning the MP2RAGE model is not invertible at this point, even if there is a well-defined range of T1 values at $S_{1,2}$ and $S_{1,3}$ (Fig. 1). Here, we propose a MAP approach to T1 mapping that allows for generalizing MP2RAGE T1 mapping to MP3RAGE T1 mapping (Fig. 2). The MAP approach also allows for uncertainty estimation with both MP2RAGE and MP3RAGE. This maximum a posteriori method expands upon the original point estimate T1 mapping introduced by Marques et al. [10]. While our method allows for T1 mapping with a third GRE signal, the third GRE signal is not necessary for this approach.

### 2.2.1. Monte Carlo simulation

To estimate MAP T1, we perform a Monte Carlo simulation, calculating the values of the gradient echo readouts (GREs) based on the acquisition parameters and the MP2RAGE model [10], adding Gaussian noise based on a noise level estimate, and using these calculated GRE values to find the corresponding values of the T1-weighted image $S_{1,2}$ and/or $S_{1,3}$ for both MP2RAGE and MP3RAGE. We refer to these methods as MAP MP2RAGE and MAP MP3RAGE to contrast with the original point estimate MP2RAGE introduced by Marques et al. [10]. We also refer to the MAP estimate of T1 as a T1 estimate to differentiate from T1 values generated by inverting a forward model as in MP2RAGE with Marques et al. [10] or from curve fitting with multi-TI IR as in Cronin et al. [20]. While our acquisition consisted of three GREs, we also investigated point estimate MP2RAGE and MAP MP2RAGE by using the first two GREs alone.

Noise in complex-valued MRI acquisitions is well-modeled by a Gaussian for the real and imaginary parts [30]. To measure the estimated noise level, we computed the standard deviation of the values in the corpus callosum of all acquired GREs from an example subject. We manually selected an ellipsoidal region of interest in the splenium of the corpus callosum. We chose the largest standard deviation across all real and imaginary parts of the acquired GREs and scaled it to have approximately the same contrast-to-noise ratio for the range of calculated GRE signals using the MP2RAGE model. We used the same value of noise in quadrature for the real and imaginary parts of all the GREs for the Monte Carlo simulation. We ran the Monte Carlo simulation for 10 million trials for a range of linearly spaced B1 correction factors ranging from 0 to 2.

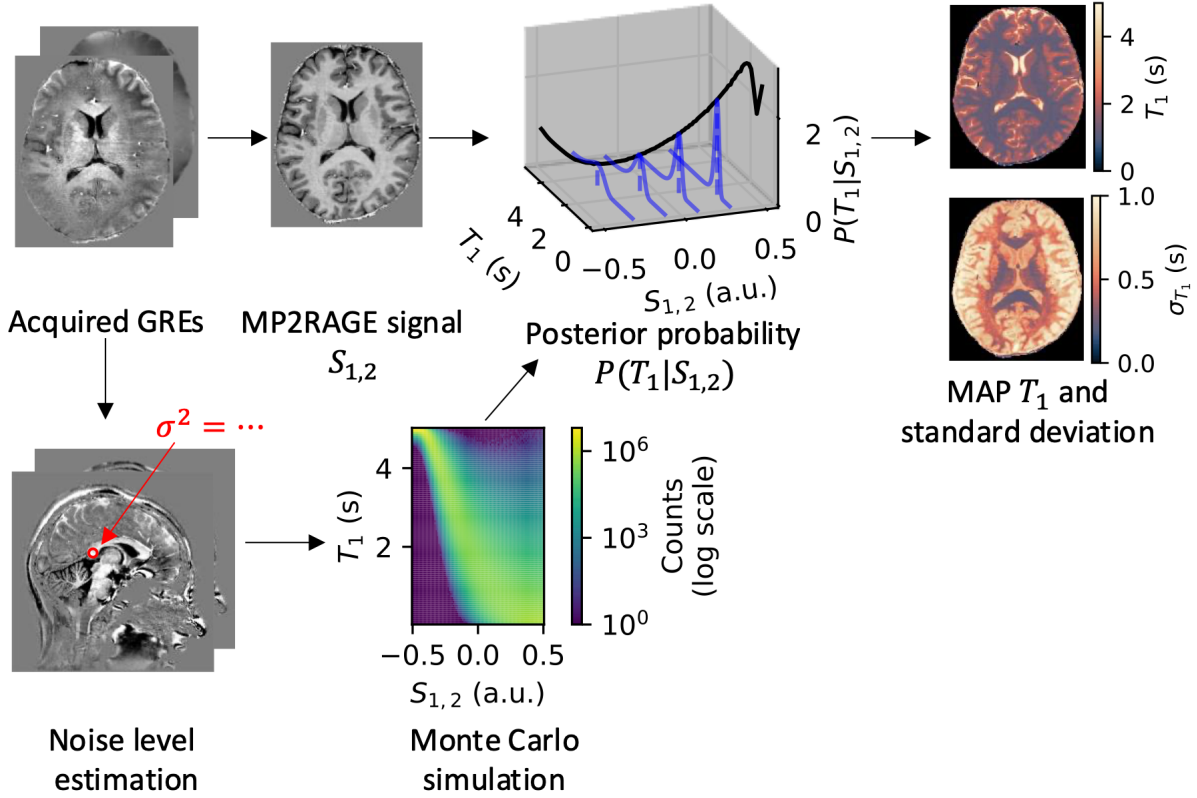

**Fig. 3.** Generating our MAP T1 estimate from an MP2RAGE image involves creating a posterior distribution $P(T_1|S_{1,2})$. We measured the noise level in the corpus callosum of the acquired GREs and ran a Monte Carlo simulation to find the values of the MP2RAGE T1-weighted image $S_{1,2}$ from the GREs with additive Gaussian noise. We normalized the counts generated across all values of T1 for each value of $S_{1,2}$ to form the posterior distribution, from which we calculated the MAP estimate of T1 and the standard deviation of T1, $\sigma_{T_1}$.

By measuring noise from a region of interest in the corpus callosum, we assumed the noise level is homogeneous across the entire image. In fact, the noise level is locally varying. We could have instead found a local estimate of noise using the model residuals from the three GRE signals [31]. However, a full characterization of the MP2RAGE model fit given nonlinearities and artifacts present in the images would require a deeper exploration, potentially considering more than three GRE signals to provide a robust estimate of noise level. We aim to describe biases present in the MP2RAGE model, and a local estimation of the noise level in this way assumes that the model is fit appropriately. Therefore, while the assumption of a homogeneous noise level is likely violated, we treat the noise level as a fixed parameter and instead explore the impact of varying noise levels in Sec. 3.3. We selected a region of interest in the splenium of the corpus callosum to measure noise, as the signal in this region is fairly homogeneous, but it is likely to be an overestimate due to the decrease in coil sensitivity with distance from the receive coil array.

For the Monte Carlo simulation, we calculated what $S_{1,2}$ would be given a uniform distribution of T1 values from 0 to 5 and a B1 correction factor. We binned the counts of $S_{1,2}$ into 100 bins from -0.5 to 0.5. The posterior distribution $P(T_1|S_{1,2})$ can be calculated using Bayes' theorem as

$$P(T_1|S_{1,2}) = \frac{P(S_{1,2}|T_1)P(T_1)}{P(S_{1,2})},  \tag{2}$$

where the likelihood $P(S_{1,2}|T_1)$ is calculated by normalizing the counts in a bin $S_{1,2}$ for a given T1 value by the total number of counts for a given T1 value, $P(T_1)$ is a uniform distribution from $T_1 = 0$ s to $T_1 = 5$ s, and $P(S_{1,2})$ is a constant scaling factor calculated such that $P(T_1|S_{1,2})$ integrates to 1.

From this distribution, we can calculate any statistical metric of interest. Here, we calculated the MAP estimate of T1, $T_{1_{MAP}} = \underset{T_1}{\operatorname{argmax}} P(T_1|S_{1,2})$. We calculated the standard deviation of T1 to serve as an uncertainty estimate (Fig. 3). To extend this method to multiple images with MAP MP3RAGE, we ran the Monte Carlo simulation for three inversion times and normalized the counts across pairs of T1-weighted images $S_{1,2}, S_{1,3}$, and so on. We calculated metrics for the joint distribution $P(T_1|S_{1,2}, S_{1,3})$ using information from all three acquired GREs.

We compared the original point estimate MP2RAGE T1 maps to the MAP MP2RAGE and MAP MP3RAGE T1 maps, registered to multi-TI IR space. We calculated the relative value of the MAP MP2RAGE and MAP MP3RAGE T1 maps to the original point estimate MP2RAGE T1 maps for quantitative comparison.

### 2.3. Multi-TI IR T1 mapping

The multi-TI IR T1 maps were created using a voxelwise nonlinear least squares fit. Similar to the acquisition protocol described in [8], the multi-TI IR acquisition was SIR, consisting of 14 samples along the inversion recovery curve at inversion times $t_i = 6$ ms, 10 ms, 16 ms, 26 ms, 43 ms, 68 ms, 110 ms, 178 ms, 288 ms, 468 ms, 760 ms, 1230 ms, 2000 ms, and 8000 ms with a constant pre-delay time of $t_d = 2.5$ ms. We acquired these images for five axial slices with 2 mm thickness in each patient, and the scan time was approximately 7 minutes. We fit the T1 maps as in [32], assuming the relaxation rates of a two-pool model with free water and macromolecular protons are equal, $R_{1f} \approx R_{1m}$. For B1 correction of the multi-TI IR images, we simulated the inversion pulse to find the inversion pulse coefficient $S_m$ over a range of $B_1^+$ values [32]. We fit the data to the biexponential SIR model using a nonlinear least squares fitting algorithm. We use the reciprocal of the longitudinal relaxation rate of free water as our value for T1; that is, $T_1 = 1/R_{1f}$. Note that the observed T1 is only equal to the relaxation time of free water because we assumed that the relaxation rates of the bound pool and free pool are equivalent. When this assumption does not hold, the observed T1 may be different [33].

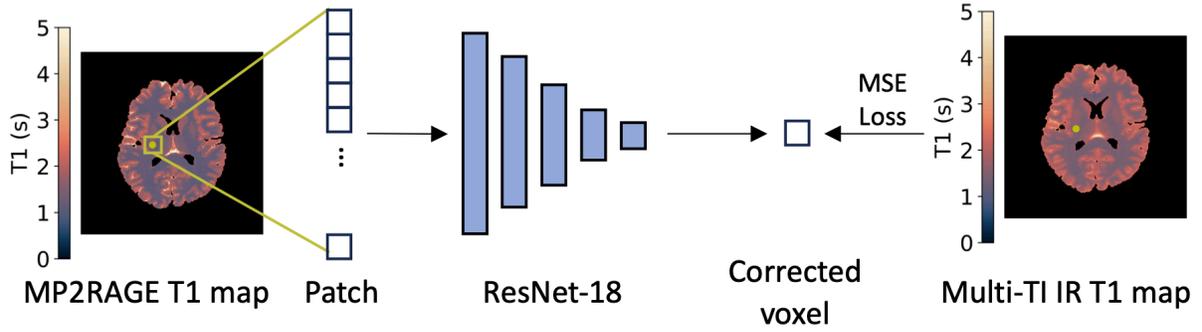

**Fig. 4.** Calibrating the MP2RAGE T1 map to the multi-TI IR T1 map involves inputting a patch of voxels from the MP2RAGE T1 map into a ResNet-18 model that outputs a single corrected voxel estimate, which we compare to the multi-TI IR T1 value at the center of the patch using mean square error (MSE) loss.

We compared the original point estimate MP2RAGE T1 maps, the MAP MP2RAGE/MAP MP3RAGE T1 maps, and the multi-TI IR T1 maps, all registered to multi-TI IR space. We compared the relative value compared to multi-TI IR T1 values and calculated the root mean square error (RMSE) for the WM, SGM, and CGM for quantitative comparison. We performed a Bland-Altman-style analysis where we compared the voxelwise error for each method to the multi-TI IR value to describe the bias and variance for each tissue type. Mean square error is a sum of bias squared and variance. The Bland-Altman-style analysis allows us to separately examine bias, resulting from systematic differences in the models, and variance, which results from noise and variability inherent in the image acquisition.

### 2.4. Calibration of MAP MP2RAGE T1 estimates to multi-TI IR T1 values

To correct differences between the MAP MP2RAGE T1 maps and multi-TI IR T1 maps, we applied a patch-based ResNet-18 [34]. The patch-based ResNet-18 is a convolutional neural network that we can use as a data-driven calibration model. We input a patch of MAP MP2RAGE quantitative T1 estimates to provide spatial context to the network, and we train the model to output the corresponding multi-TI IR quantitative T1 value at the center voxel of the patch.

We used leave-one-out cross-validation across the four subjects, training on two subjects, validating on one subject, and testing on the final subject across four folds. The input to the network is a $5 \times 5$ patch from an axial slab of the MAP MP2RAGE T1 map, and the output is the multi-TI IR-corrected MAP MP2RAGE T1 estimate of the center voxel in the patch (Fig. 4). We

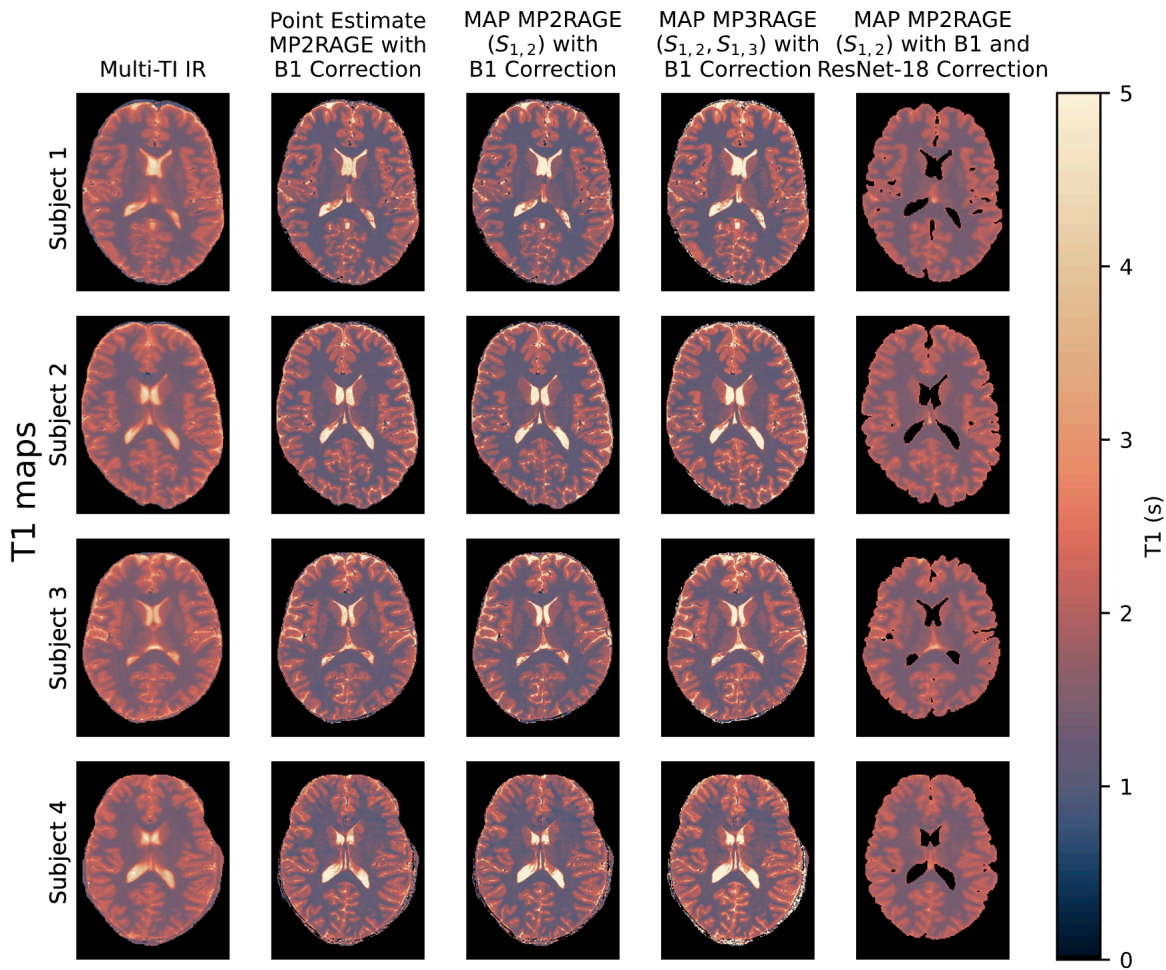

**Fig. 5.** Comparing the point estimate MP2RAGE T1 maps with the MAP MP2RAGE and MAP MP3RAGE T1 maps, the values are very similar. However, there is a bias between MAP MP2RAGE T1 estimates and multi-TI IR T1 values that the ResNet-18 calibration model reduces. See Fig. 6 and Fig. 7 for a discussion of the relative values. Note the ResNet-18 calibration model outputs voxels in white and gray matter only.

chose to use 2D in-plane (i.e., axial) patches for the calibration model. 3D volumetric patches would require either zero-padding patches that fall along the edges of the volume (and therefore assuming there is no anatomy outside of the superior-inferior field of view of the multi-TI IR acquisition), or they would only allow us to calibrate voxels whose corresponding volumetric patch falls entirely within the field of view (which would reduce the number of axial slices from the five we collected). We used patches whose center lies in the brain tissue (not CSF or background), using mean square error loss between the output voxel value and the multi-TI IR voxel value. The learning rate was $10^{-5}$, with batches of 256 patches for 10,000 steps. We checked the validation loss every 50 steps and stopped early if the validation loss has not decreased in 1,000 steps. To determine if uncertainty information assisted in calibrating the MAP

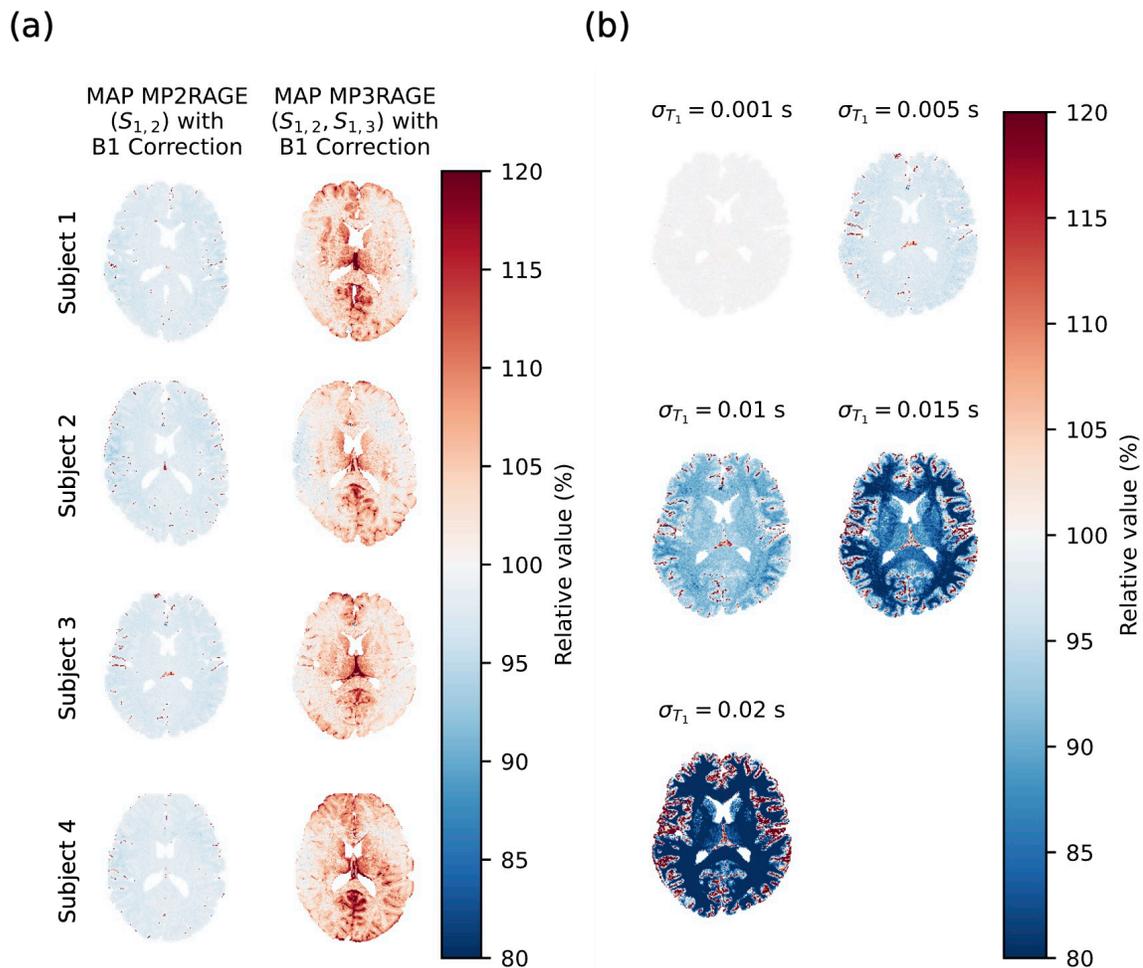

**Fig. 6.** The relative value of the T1 estimates from MAP MP2RAGE and MAP MP3RAGE compared to the original point estimate MP2RAGE from Marques et al. shows similar values for MAP MP2RAGE. The T1 estimates from MAP MP3RAGE are more sensitive to B1 inhomogeneities (a). Across noise levels, MAP MP2RAGE T1 estimates decrease in the white matter and subcortical gray matter and increase in cortical gray matter relative to point estimate MP2RAGE, with smaller noise levels resulting in more similar T1 estimates compared to point estimate MP2RAGE (b).

MP2RAGE T1 estimates, we trained a model with the MAP MP2RAGE T1 patches alone as well as a second model with the standard deviation of T1 as a second channel for the input image.

Next, we tested the sensitivity of our calibration network to the patch size and the noise level estimate. We investigated the impact of patch size on calibration performance using by training separate models on patch sizes of 1×1, 5×5, 9×9, and 13×13 with a constant noise level standard deviation of 0.005 s. To explore the impact of estimated noise level on the model's performance, we trained separate models on MAP MP2RAGE T1 maps generated using noise level standard deviations of 0.001 s, 0.005 s, 0.01 s, 0.015 s, and 0.02 s with a constant patch size of 5×5. For the Monte Carlo simulations in the noise level parameter sweep, we ran 1 million trials. The hyperparameters for training the models for the sensitivity analysis were the same as above.

We compared the mean RMSE across folds of the MAP MP2RAGE T1 maps with B1 and ResNet-18 correction to the mean RMSE across folds for the MAP MP2RAGE T1 with B1 correction alone. We investigated the performance of the network with the T1 patch alone as well as with the T1 patch with the standard deviation channel. We validate the performance of the network by comparing the average RMSE of the test subject's images across folds using a paired $t$-test with a significance level of $\alpha = 0.05$.

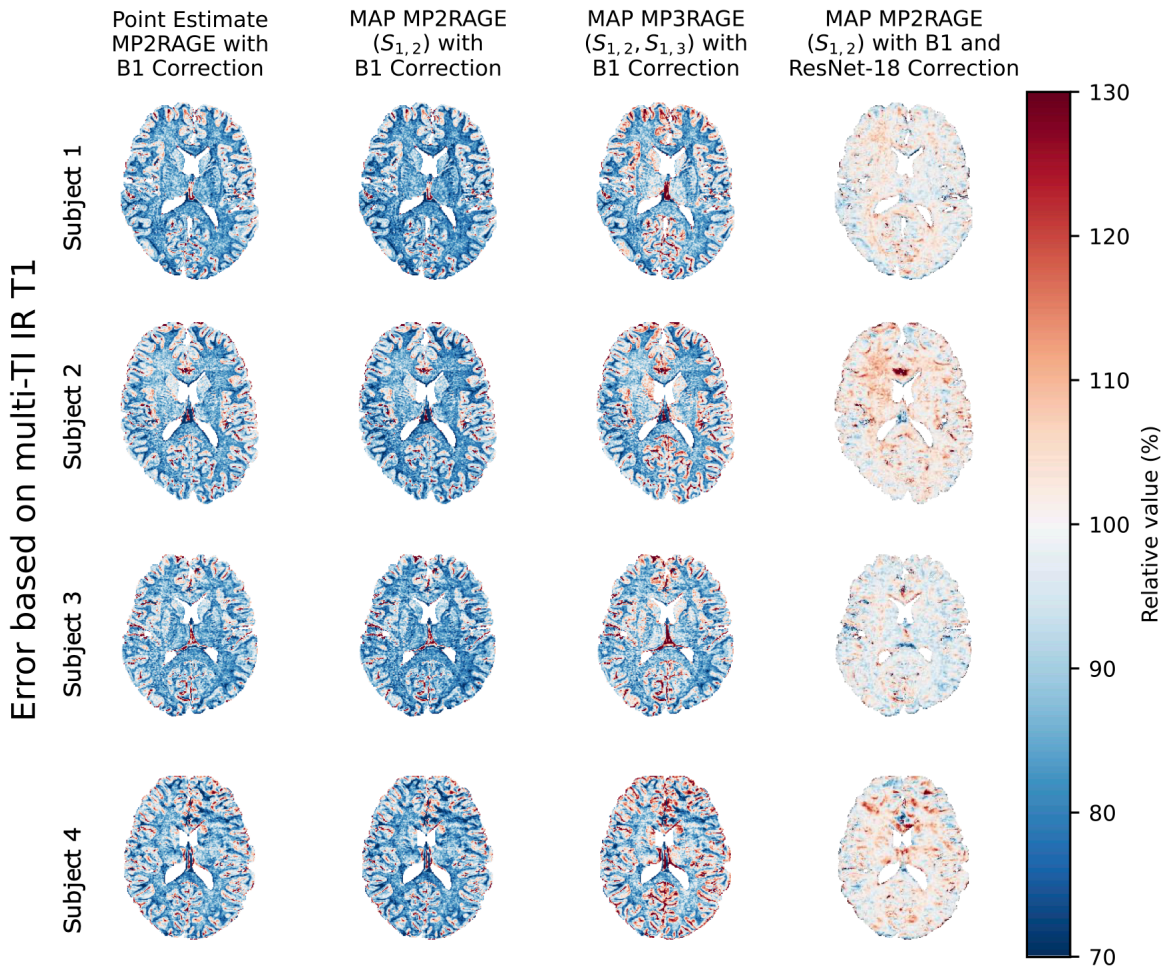

**Fig. 7.** There is a substantial error for the point estimate MP2RAGE, MAP MP2RAGE, and MAP MP3RAGE T1 maps relative to the multi-TI IR T1 maps. The average relative value across subjects for MAP MP2RAGE compared to multi-TI IR is 95.3% in cortical gray matter, 89.1% in subcortical gray matter, and 82.8% in white matter. The patch-based ResNet-18 calibration model reduces the error (average relative value of 99.9% in cortical gray matter, 100.7% in subcortical gray matter, and 101.6% in white matter).

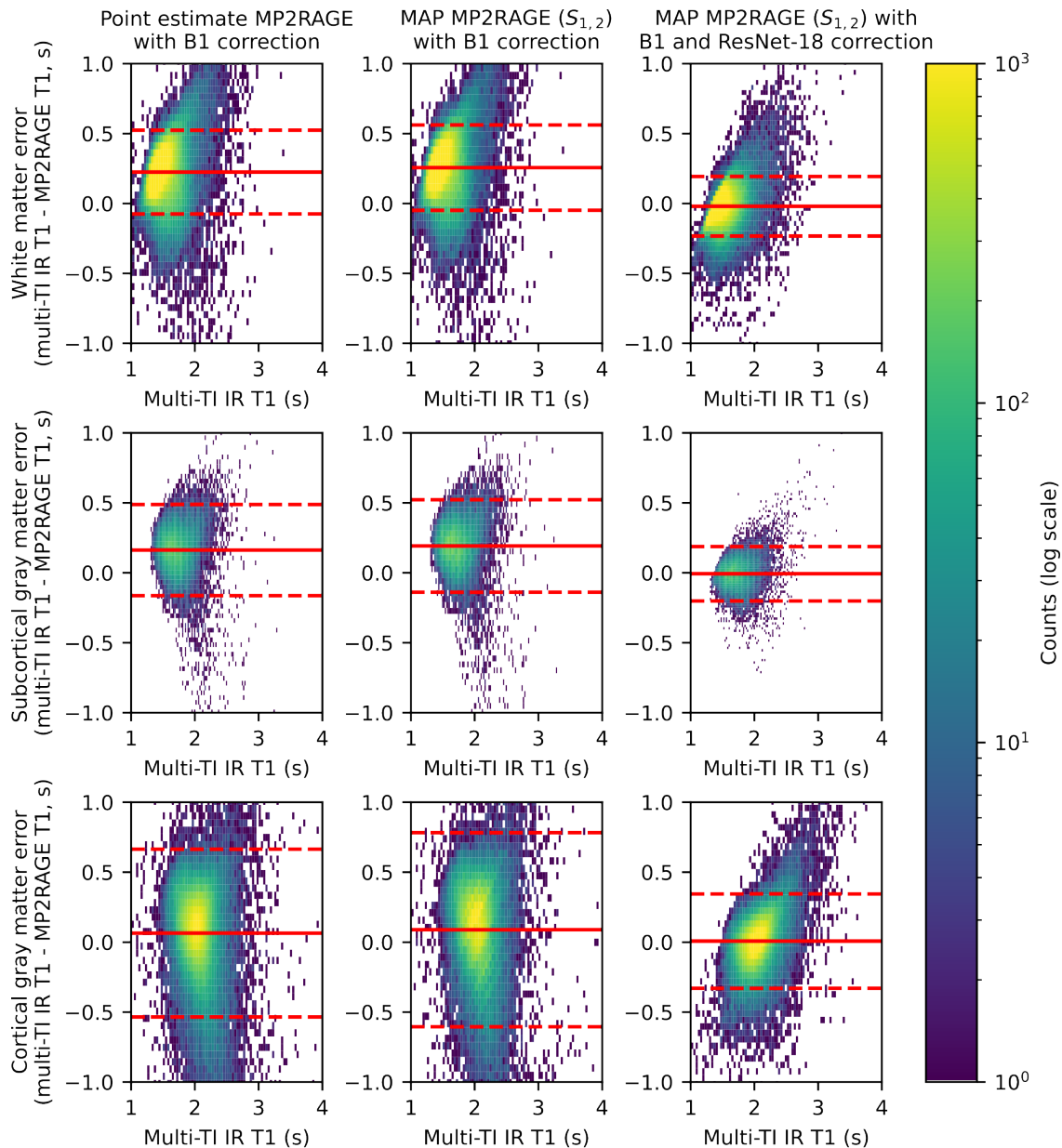

**Fig. 8.** Bland-Altman-style comparison between the MP2RAGE T1 maps and multi-TI IR T1 maps demonstrates high variance in the error for the original point estimate MP2RAGE and MAP MP2RAGE with B1 correction, as well as a bias in all tissue types. The patch-based ResNet-18 calibration model reduces the bias and moderately reduces the variance, though the cortical gray matter variance remains high after calibration. Solid lines indicate mean error, while dashed lines indicate plus/minus 1.96 times the standard deviation.

## 3. Results

### 3.1. Comparison of MAP MP2RAGE and MAP MP3RAGE T1 estimates to point estimate MP2RAGE T1 values

The original point estimate MP2RAGE T1 values with B1 correction are very similar to the MAP MP2RAGE T1 estimates (Fig. 5). MAP MP2RAGE T1 estimates are very similar to point estimate MP2RAGE T1 values. However, T1 estimates from MAP MP3RAGE are overestimated compared to point estimate MP2RAGE, and the overestimation varies across the

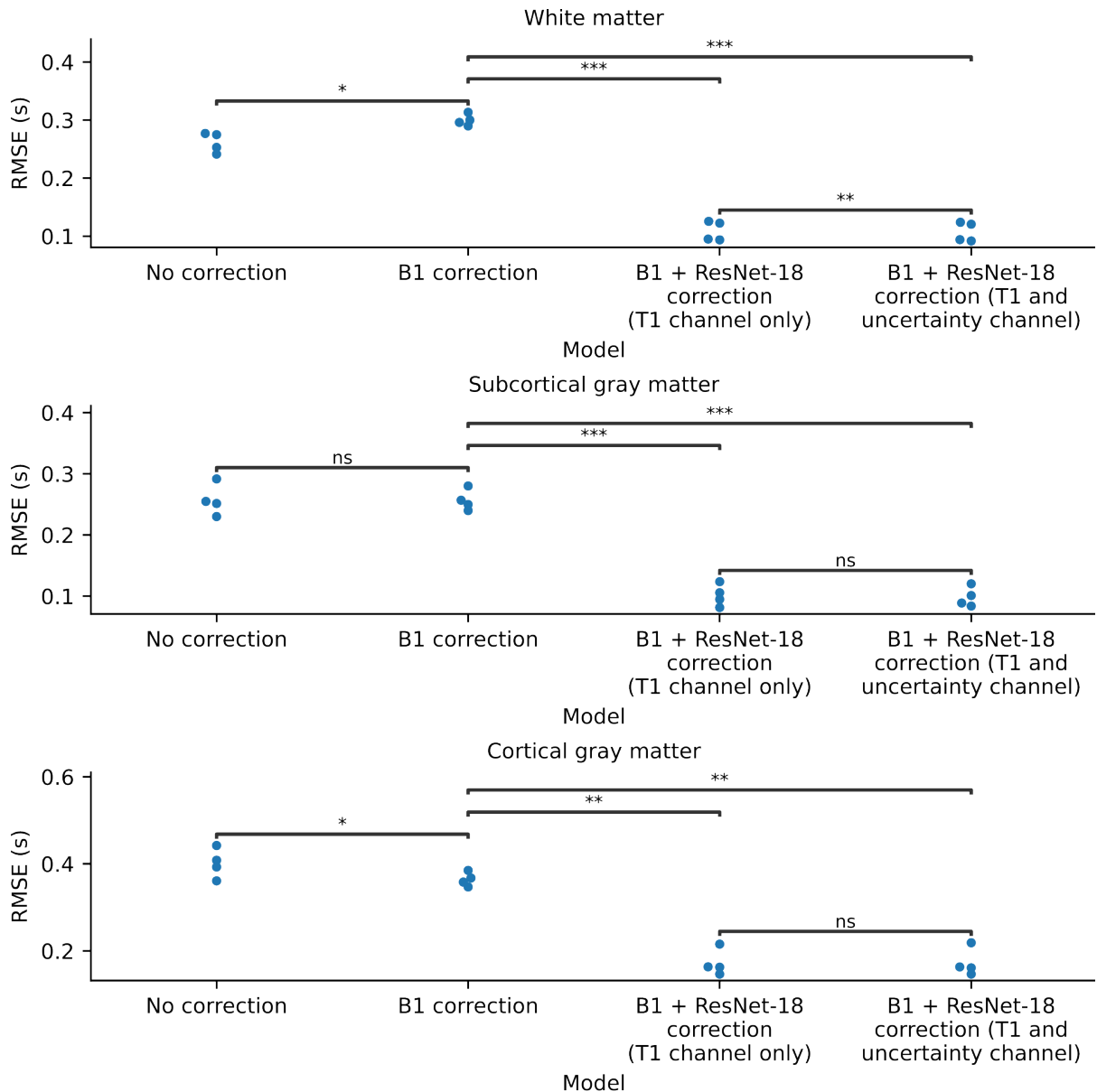

**Fig. 9.** Using B1 correction alone results in a significant change in RMSE in the white matter and cortical gray matter, but not in the subcortical gray matter. The calibration ResNet-18 model significantly reduces the tissue-dependent bias compared to B1 correction alone in all tissue types. The addition of the uncertainty estimate as a second channel to the network does not substantially change the RMSE across the four folds for all tissue types except white matter. *: $p < 0.05$, **: $p < 0.01$, ***: $p < 0.001$, ns: not significant using a paired $t$-test.

**Table 1.** T1 values across methods and RMSE compared to multi-TI IR T1 values.

| Method | Tissue Type* | Average T1 value (s) | Average RMSE (s) |
|---|---|---|---|
| Multi-TI IR (with B1 correction) | WM | 1.48 ± 0.03 | -- |
| | SGM | 1.74 ± 0.05 | -- |
| | CGM | 2.05 ± 0.03 | -- |
| MP2RAGE (no B1 correction) | WM | 1.28 ± 0.04 | 0.26 ± 0.02 |
| | SGM | 1.57 ± 0.08 | 0.26 ± 0.03 |
| | CGM | 2.10 ± 0.06 | 0.40 ± 0.03 |
| MP2RAGE (with B1 correction) | WM | 1.22 ± 0.04 | 0.30 ± 0.01 |
| | SGM | 1.55 ± 0.07 | 0.26 ± 0.02 |
| | CGM | 1.96 ± 0.05 | 0.36 ± 0.02 |
| MP2RAGE (B1 + ResNet-18 correction, T1 channel only) | WM | 1.50 ± 0.03 | 0.11 ± 0.02 |
| | SGM | 1.75 ± 0.06 | 0.10 ± 0.02 |
| | CGM | 2.04 ± 0.03 | 0.17 ± 0.03 |
| MP2RAGE (B1 + ResNet-18 correction, T1 and uncertainty channel) | WM | 1.50 ± 0.03 | 0.11 ± 0.02 |
| | SGM | 1.74 ± 0.06 | 0.10 ± 0.02 |
| | CGM | 2.04 ± 0.03 | 0.17 ± 0.03 |

*WM: white matter, SGM: subcortical gray matter, CGM: cortical gray matter

image (Fig. 6). The MAP MP2RAGE T1 estimates also depend substantially on noise level, with lower noise levels agreeing more closely with point estimate MP2RAGE. With our selected noise level of 0.005 s, the average relative value of MAP MP2RAGE T1 estimates compared to point estimate MP2RAGE T1 values is 98.5% in cortical gray matter, 98.1% in subcortical gray matter, and 97.4% in white matter (Fig. 7).

### 3.2. Comparison to multi-TI IR T1 values

Though the point estimate MP2RAGE, MAP MP2RAGE, and MAP MP3RAGE T1 maps are all comparable, they all have a substantially lower value compared to multi-TI IR T1 maps, with white matter the lowest with an average relative value of 82.8% (Fig. 7). A Bland-Altman-style comparison plotting the error between multi-TI IR T1 values and MAP MP2RAGE/MAP MP3RAGE estimates versus multi-TI IR T1 values demonstrates for MAP MP2RAGE and point estimate MP2RAGE, both methods tend to underestimate T1 compared to multi-TI IR (Fig. 8). There is a very large variance in the error in both tissue types, with a slightly lower variance in the WM and SGM than in the CGM.

### 3.3. Calibration of MAP MP2RAGE T1 estimates to multi-TI IR T1 values

The patch-based ResNet-18 calibration model reduces the magnitude of the error for each of the subjects, with a relative value near 100% for all tissues (Fig. 7). The Bland-Altman-style comparison shows that the patch-based ResNet-18 calibration network reduces the magnitude of the bias in both tissue types. Additionally, the network reduces the variance of the error in both tissue types (Fig. 8). The calibration ResNet-18 significantly reduces the RMSE in the WM, SGM and CGM, both using the T1 channel alone as well as using the T1 channel and standard deviation channel (Fig. 9). Table 1 provides the mean T1 value and standard deviation of T1 across subjects for each method, as well as the mean RMSE and standard deviation of RMSE across folds compared to multi-TI IR T1 values. For the patch-based ResNet-18, the addition of the standard deviation as a second channel for the input patches to the network does not substantially change the RMSE compared to the T1 channel alone.

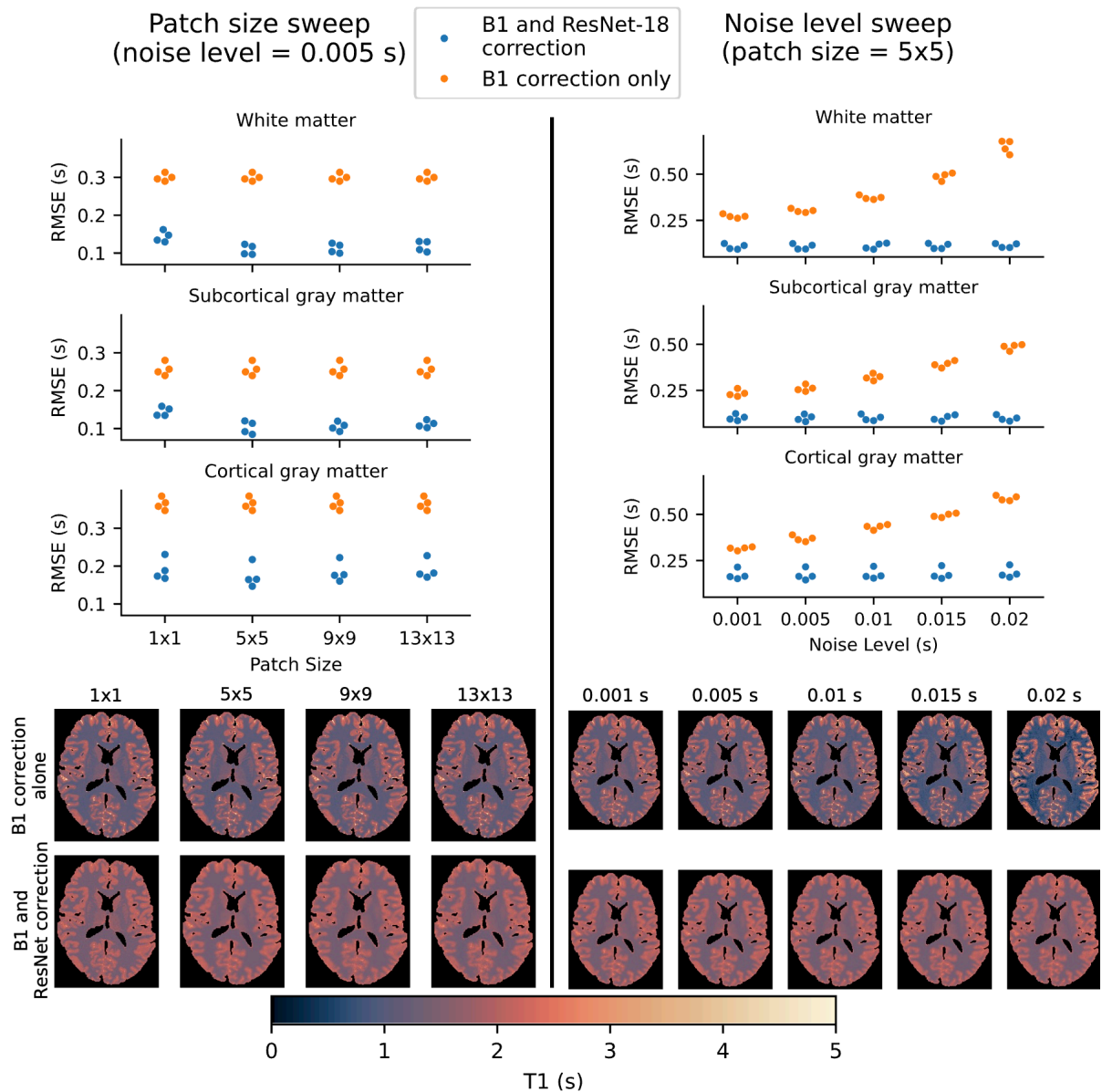

**Fig. 10.** The ResNet-18 calibration model's performance depends on the amount of spatial context provided with the patch size, with 5×5 performing slightly better than the other patch sizes. The RMSE produced by MAP MP2RAGE T1 maps increases with higher standard deviations of noise for the Monte Carlo simulation, but the calibration model corrects this error to a relatively constant value across different values of the noise level, albeit with a higher variance in the RMSE.

The calibration network's performance depends slightly on patch size, with a patch size of 5×5 performing the best (Fig. 10). The RMSE of the MP2RAGE MAP T1 maps depends heavily on the standard deviation of the noise level selected for the Monte Carlo simulation, with higher values of the noise level resulting in higher RMSE. However, the calibration network achieves a similar corrected RMSE across all values of the noise level.

## 4. Discussion

While MAP MP2RAGE T1 estimates are similar to point estimate MP2RAGE T1 values, the MAP MP3RAGE T1 estimates are more sensitive to B1 inhomogeneities, despite B1 correction. We found a similar sensitivity to B1 inhomogeneities when comparing MAP MP2RAGE using the first and third GRE, which suggests the sensitivity may be due to the longer inversion time. The inversion times for the third GRE is not optimized to reduce B1 inhomogeneities, unlike the inversion times for the first two GREs with MP2RAGE [10].

There are substantial differences between MAP MP2RAGE T1 estimates and multi-TI IR T1 values, but we found that the error between the methods can be significantly reduced with a patch-based neural network trained on limited subjects. The MAP approach allows us to generate any statistical metric of interest, but adding the standard deviation as an uncertainty channel for the calibration network did not substantially reduce the error. The sensitivity analysis reveals that the calibration network performs best with some spatial context, as a voxelwise correction does not perform as well as patch-based corrections. We found a low noise level works best for the Monte Carlo simulation.

Previous studies have taken advantage of the long $MP2RAGE_{TR}$ to collect a third GRE but estimated T1 maps by either averaging the T1 maps from the multiple T1-weighted images as in Sun et al. [15], least-squares fitting between the two images as in Hung et al. [18], or performing T1 mapping using only one of the T1-weighted images as in Metere et al. and Caan et al. [13,14]. Our proposed MAP MP3RAGE allows for T1 mapping with multiple GREs simultaneously. A different approach introduced by Olsson et al. used a third GRE with a high flip angle to create closed-form solutions for the effective longitudinal relaxation time $T_1^*$ (accounting for partial saturation) and therefore T1 alongside $B_1^+$ [17].

While this study demonstrated the feasibility of calibrating quantitative T1 methods through a patch-based neural network, the use of a neural network for calibrating the T1 values does not explain why there are differences between the methods. Rioux et al. proposed that the difference may be due to the biexponential model for multi-TI IR, while point estimate MP2RAGE has limited samples and must use a monoexponential model that cannot account for magnetization transfer effects [21]. Multi-TI IR T1 values are corrected for magnetization transfer effects. Rowley et al. found a bias in point estimate MP2RAGE quantitative T1 values compared to variable flip angle (VFA) T1 values at 3T in vivo and in simulations that was attributable to magnetization transfer effects [35]. They found that the differences in point estimate MP2RAGE and VFA T1 mapping are not present in phantoms and concluded that these biases result from magnetization transfer effects that would not affect a homogeneous solution in a phantom [35]. Soustelle et al. also found magnetization transfer effects in point estimate MP2RAGE quantitative T1 values at 3T that are present in vivo and in simulations but are not present in homogeneous solutions in phantoms. The simulations demonstrated a bias in quantitative T1 values at 3T with a relative variation of approximately 5% to 20% depending on tissue type and inversion pulse efficiency [36]. They proposed prospective solutions that can correct for the magnetization transfer effects by collecting an additional magnetization transfer-prepared spoiled gradient-recalled acquisition and fitting with a two-pool model, reducing the bias in T1 values substantially [36,37]. Our proposed calibration model is retrospective and can be applied to data where additional acquisitions are not available for fitting with Soustelle et al.'s method, and instead we correct based on paired data collected from a few subjects.

Previous work by Bottomley et al. in 1987 aggregated reported T1 values from over 200 studies and fit the T1 values to a model of the form $T_1 = A\nu^B$, where $\nu$ is the Larmor frequency and $A$ and $B$ are tissue-dependent and pathology-dependent constants [38]. Using their model for normal tissue at a Larmor frequency of 300 MHz for 7T, the white matter should fall within 1.12 s to 1.58 s, and gray matter should fall within 1.23 s to 1.73 s, which overlap with our values. With B1 correction alone, our MP2RAGE quantitative T1 values agree with those reported by Rooney et al. at 7T using a modified Look-Locker IR sequence. However, our multi-TI IR T1 values and ResNet-18-corrected MP2RAGE T1 values are overestimated compared to Rooney et al. A potential source of difference is that Rooney et al. fit to a monoexponential model that does not account for magnetization transfer effects, similar to MP2RAGE [19].

This study is limited by the small sample size ($n = 4$). While we demonstrated calibration of the MAP MP2RAGE quantitative T1 estimates using leave-one-out cross-validation, we did not demonstrate the generalization of the calibration model beyond a single subject left out from training for each fold. It is likely that the learned calibration model would not generalize to new hardware or different sites. However, we showed that the calibration model can be trained using limited paired data for retrospective study. Additionally, there is a loss of image resolution when we registered from the MP2RAGE/MP3RAGE acquisition (0.7 mm × 0.43 mm × 0.43 mm) to the multi-TI IR acquisition (0.86 mm × 0.86 mm × 2 mm). We accounted for partial volume effects from the cerebrospinal fluid by eroding the cortical gray matter labels, though there may still be effects in the subcortical tissue. For the multi-TI IR model, we assumed the relaxation rates of the free and bound pools are equivalent so that the observed T1 is the same as the relaxation time of the free pool. Helms and Hagberg discussed how this assumption may not hold and can result in different T1 values [33].

The advantage of MP2RAGE as a fast method for high-resolution quantitative T1 mapping is limited by the need for B1 correction with an additional $B_1^+$ map. We used multi-TI IR with a selective IR pulse for ground truth T1 values because the model accounts for magnetization transfer effects using a biexponential recovery. We assumed the noise level is global, but we could have explored estimating a locally varying noise field from the MP3RAGE model [31]. We also assumed the $B_1^+$ field was the same along all axial slices of the multi-TI IR acquisition.

After our calibration network, the mean differences between quantitative T1 values between the methods are similar in magnitude to differences reported between different quantitative T1 mapping methods at 7T in the literature [8,10,39–41]. Note that there is always noise present in the image acquisition process, meaning the value of RMSE will always be greater than zero. The model demonstrated here allows for retrospective calibration of comparisons across sequences, but prospective studies should focus on implementing sequences for quantitative T1 mapping that account for magnetization transfer effects.

## 5. Conclusion

Here, we demonstrated a method for generating a posterior distribution of T1, allowing for T1 mapping across both MP2RAGE and MP3RAGE and generation of statistical metrics like the standard deviation of T1, and we found no substantial bias compared to typical point estimate T1 mapping with MAP MP2RAGE, though MAP MP3RAGE was sensitive to B1 inhomogeneities. We compared T1 maps from typical point estimate MP2RAGE, MAP MP2RAGE, and MAP MP3RAGE and found a substantial tissue-dependent bias compared to gold-standard multi-TI IR T1 maps, potentially a result of the monoexponential recovery assumed in the MP2RAGE model that does not account for magnetization transfer effects. To address the spatially dependent differences between the methods, we demonstrated that a patch-based ResNet-18 trained on registered MAP MP2RAGE and multi-TI IR T1 maps can reduce the error between the two methods by decreasing the magnitude of the bias and reducing the variance of the error. This method allows for combining quantitative MRI across multiple acquisitions where the parameters of interest were collected under different acquisitions. A neural network can retrospectively calibrate quantitative parameters using limited paired training data.

### Declaration of generative AI and AI-assisted technologies in the writing process

During the preparation of this work the author(s) used GitHub Copilot in order to create code segments based on task descriptions, as well as to debug, edit, and autocomplete code, and ChatGPT to assist in structuring sentences and performing grammatical checks. After using this tool/service, the authors reviewed and edited the content as needed and take full responsibility for the content of the publication.

**Supplementary Materials**

Quantitative T1 mapping is sensitive to acquisition parameters and the B1 correction factor (Supplementary Fig. 1). The lookup table for point estimate MP2RAGE demonstrates this sensitivity as well (Supplementary Fig. 2).

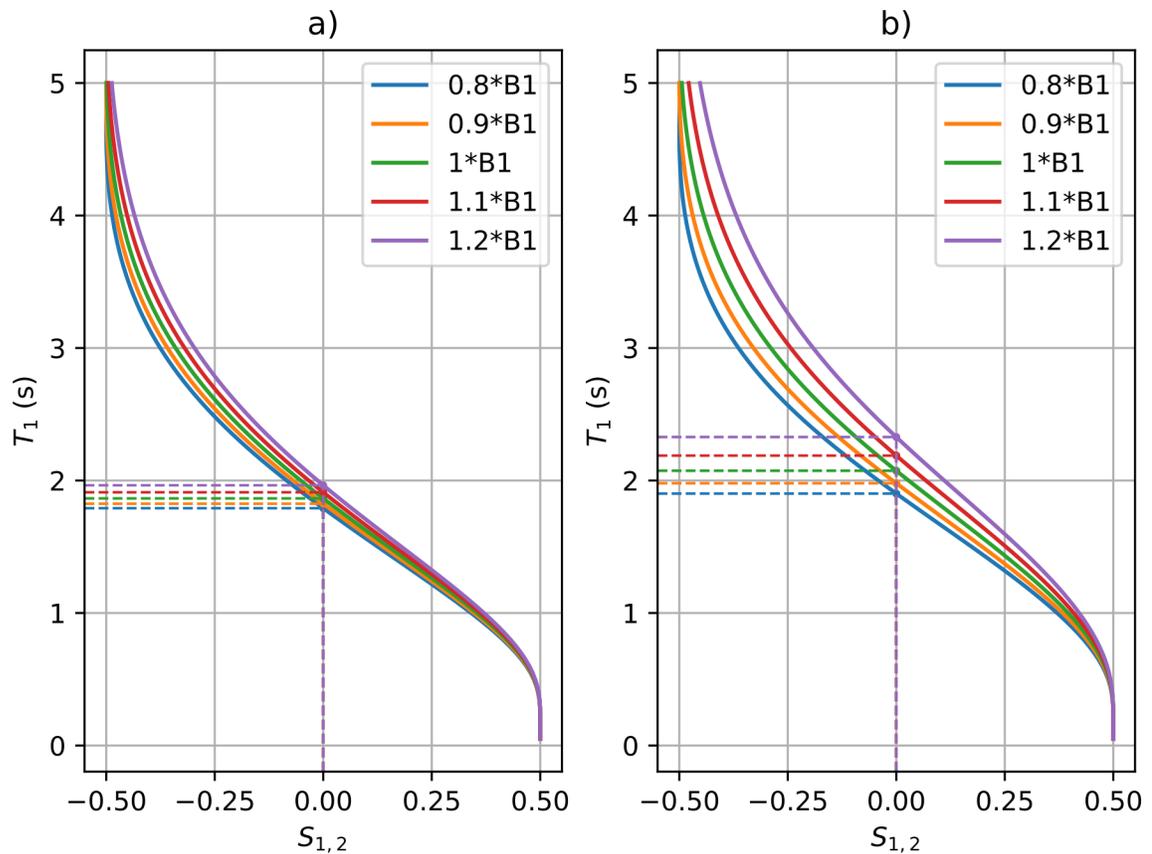

**Supplementary Fig. 1.** MP2RAGE T1 values are sensitive to the values of the acquisition parameters, as well as the B1 correction factor. Acquisition parameters are $\text{MP2RAGE}_{TR} = 8.25$ s, $TR = 6$ ms, $TI_1 = 1010$ ms, $TI_2 = 3310$ ms, with 225 excitation pulses, flip angles of 4 degrees, and an inversion pulse efficiency of 0.84 for a) as in this study, and $\text{MP2RAGE}_{TR} = 8.5$ s, $TR = 6.9$ ms, $TI_1 = 1000$ ms, $TI_2 = 3000$ ms, with 252 excitation pulses, flip angles of 5°, and an inversion pulse efficiency of 0.84 for b) as in a previous study by Choi et al. [1].

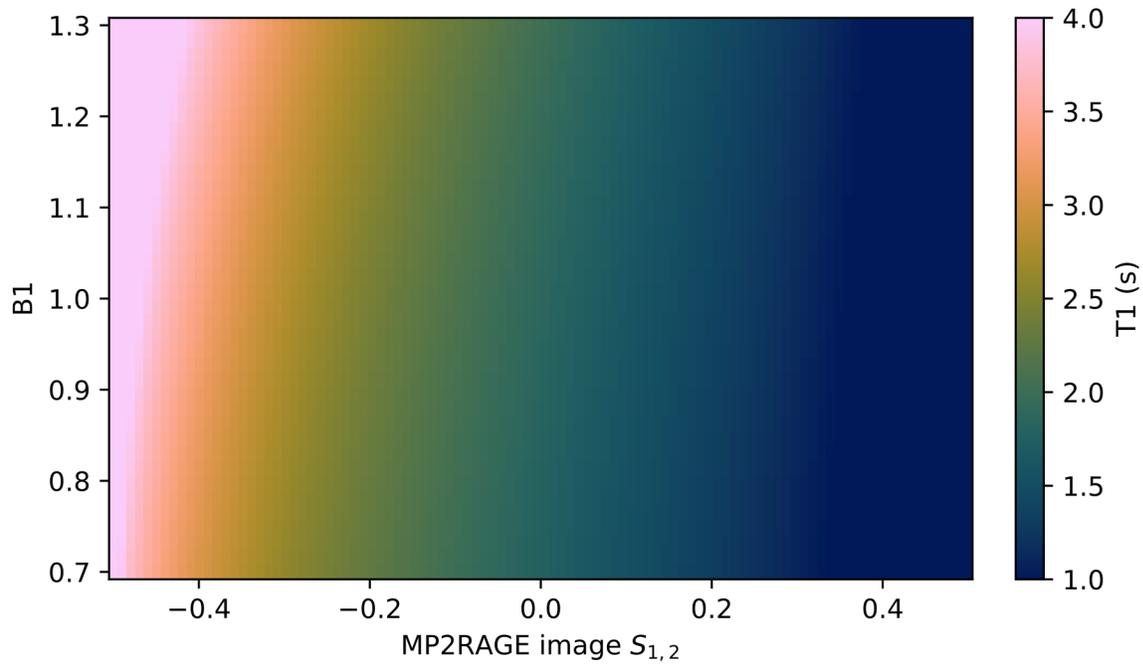

**Supplementary Fig. 2.** The lookup table for T1 from point estimate MP2RAGE T1 mapping demonstrates sensitivity to the B1 correction factor. The acquisition parameters are the same as in Supplementary Fig. 1 (a).